\documentstyle[11pt,equation,epsf]{article}
\input{psfig.sty}
\begin{document}
\newcommand{\tcr}{T_{cr}}
\newcommand{\chit}{\tilde{\chi}}
\newcommand{\phit}{\tilde{\phi}}
\newcommand{\df}{\delta \phi}
\newcommand{\dr}{\delta \rho}
\newcommand{\dxl}{\delta x_{\Lambda}}
\newcommand{\dkl}{\delta \kappa_{\Lambda}}
\newcommand{\dkx}{\delta x_{\Lambda}}
\newcommand{\dk}{\delta \kappa}
\newcommand{\dlt}{\delta \tilde{\lambda}}
\newcommand{\dkg}{\delta \kappa_{G}}
\newcommand{\dkcr}{\delta \kappa_{cr}}
\newcommand{\dxg}{\delta x_{G}}
\newcommand{\dx}{\delta x}
\newcommand{\lx}{\lambda}
\newcommand{\Lx}{\Lambda}
\newcommand{\ex}{\epsilon}
\newcommand{\ks}{k_s}
\newcommand{\gb}{\bar{g}}
\newcommand{\gt}{\tilde{g}}
\newcommand{\lb}{{\bar{\lambda}}}
\newcommand{\lbz}{\bar{\lambda}_0}
\newcommand{\lt}{\tilde{\lambda}}
\newcommand{\lr}{{\lambda}_R}
\newcommand{\xr}{x_R}
\newcommand{\xt}{\tilde{x}}
\newcommand{\lrt}{{\lambda}_R(T)}
\newcommand{\lbr}{{\bar{\lambda}}_R}
\newcommand{\lk}{{\lambda}(k)}
\newcommand{\lbk}{{\bar{\lambda}}(k)}
\newcommand{\lbkt}{{\bar{\lambda}}(k,T)}
\newcommand{\ltk}{\tilde{\lambda}(k)}
\newcommand{\mx}{{m}^2}
\newcommand{\mxk}{{m}^2(k)}
\newcommand{\mxkt}{{m}^2(k,T)}
\newcommand{\mb}{{\bar{m}}}
\newcommand{\mt}{\tilde{m}}
\newcommand{\mr}{{m}^2_R}
\newcommand{\mrt}{{m}^2_R(T)}
\newcommand{\mk}{{m}^2(k)}
\newcommand{\rhoa}{\rho_1}
\newcommand{\rhob}{\rho_2}
\newcommand{\nua}{\nu_A}
\newcommand{\nub}{\nu_B}
\newcommand{\nuab}{\bar{\nu}_A}
\newcommand{\nubb}{\bar{\nu}_B}
\newcommand{\rhb}{\bar{\rho}}
\newcommand{\rht}{\tilde{\rho}}
\newcommand{\rhta}{\tilde{\rho}_1}
\newcommand{\rhtb}{\tilde{\rho}_2}
\newcommand{\rhz}{\rho_0}
\newcommand{\rhzr}{\rho_{0R}}
\newcommand{\rhza}{\rho_{10}}
\newcommand{\rhzb}{\rho_{20}}
\newcommand{\rhzt}{\tilde{\rho}_0}
\newcommand{\rhzta}{\tilde{\rho}_{10}}
\newcommand{\rhztb}{\tilde{\rho}_{20}}
\newcommand{\rhzk}{\rho_0(k)}
\newcommand{\rhzkt}{\rho_0(k,T)}
\newcommand{\kx}{\kappa}
\newcommand{\kt}{\tilde{\kappa}}
\newcommand{\kk}{\kappa(k)}
\newcommand{\ktk}{\tilde{\kappa}(k)}
\newcommand{\Gammat}{\tilde{\Gamma}}
\newcommand{\Gammak}{\Gamma_k}
\newcommand{\wt}{\tilde{w}}
\newcommand{\be}{\begin{equation}}
\newcommand{\ee}{\end{equation}}
\newcommand{\een}{\end{subequations}}
\newcommand{\ben}{\begin{subequations}}
\newcommand{\beq}{\begin{eqalignno}}
\newcommand{\eeq}{\end{eqalignno}}
\newcommand{\lsim}{\begin{array}{c}<\vspace{-0.32cm}\\\sim\end{array}}
\newcommand{\gsim}{\begin{array}{c}>\vspace{-0.32cm}\\ \sim\end{array}}
\pagestyle{empty}
\noindent
\begin{flushright}
\end{flushright} 
\vspace{2cm}
\begin{center}
{ \Large Renormalization-Group Study  \\
 of Weakly First-Order Phase Transitions} 
\\ \vspace{1.5cm}
N. Tetradis \\
\vspace{0.5cm}
CERN, Theory Division, \\ 
CH-1211, Geneva 23, Switzerland \\
\vspace{3cm}
\end{center}
\begin{abstract}{
We study the universal critical behaviour near
weakly first-order phase transitions for a three-dimensional 
model of two coupled scalar fields -- the cubic anisotropy model.
Renorma-lization-group techniques are employed within 
the formalism of the effective average action.
We calculate the universal form of the coarse-grained free energy and 
deduce the ratio of susceptibilities on either
side of the phase 
transition. We compare our results with those 
obtained through Monte Carlo simulations and the $\epsilon$-expansion.
}
\end{abstract}


\clearpage

\setlength{\textwidth}{16cm}
\pagestyle{plain}
\setcounter{page}{1}

A quantitative description of weakly first-order phase transitions 
(which are typically fluctuation driven)
is necessary for a wide range of problems. Apart from their
relevance to statistical systems
(such as superconductors or anisotropic systems, like the 
one we consider below), weakly first-order phase transitions
often appear in the study of high-temperature field theories. 
As a result they have important implications for cosmology. 
An example is provided by the electroweak phase transition, which is weakly 
first order for Higgs boson masses between approximately
40 and 80 GeV \cite{review,me}. The calculation of the
baryon number that can be generated during the electroweak phase
transition requires a precise determination of its properties.

A simple example of a system with an arbitrarily weakly 
first-order phase transition is the cubic anisotropy model
\cite{rudnick, aharony}. In field-theoretical language,  
this corresponds to 
a theory of two real scalar fields 
$\phi_a~(a=1,2)$ in three dimensions,  
invariant under the discrete symmetry
$(1 \leftrightarrow - 1,
2 \leftrightarrow - 2,
1 \leftrightarrow  2)$. 
It can also be considered as the effective description of
the four-dimensional high-temperature theory with the same
symmetry, at energy scales smaller than the temperature
\cite{transition,alford,stefan}. It is relevant for 
the phase transitions in
multi-Higgs extensions of the Standard Model.

Recently, the properties of the weakly 
first-order phase transitions in this model have been studied in detail,
within the $\epsilon$-expansion \cite{arnolde} or through
Monte Carlo simulations \cite{arnoldl}. In particular,
universal amplitudes have been computed, which describe 
the relative discontinuity
of various physical quantities along the phase transition, in the 
limit when the transition becomes arbitrarily weakly first order.
A discrepancy has been observed in the predictions for the 
universal ratio of susceptibilities $\chi_+/\chi_-$ on either
side of the phase transition obtained through
Monte Carlo simulations or the $\epsilon$-expansion
\cite{arnolds}. More specifically, the Monte Carlo simulations
predict $\chi_+/\chi_-=4.1(5)$, while the first three orders of the
$\ex$-expansion give $\chi_+/\chi_-=2.0, 2.9, 2.3$ respectively.

In this letter we present an alternative approach that employs the 
exact renormalization group \cite{wilson}.
It is based on the effective average action $\Gamma_k$ 
\cite{average}, which is a coarse-grained
free energy with an infrared cutoff. More precisely, 
$\Gamma_k$  incorporates the effects
of all fluctuations with momenta $q^2 > k^2$, but not those with
$q^2 < k^2$. In the limit
$k \rightarrow 0$, the effective average action 
becomes the standard effective action
(the generating functional of the 1PI Green functions), while at a high
momentum scale (of the order of the ultraviolet cutoff) 
$k =\Lx \rightarrow \infty$,
it equals the classical or bare
action.
It is formulated in continuous (Euclidean) space and all symmetries
of the model are preserved. 
The exact non-perturbative flow equation 
for the scale dependence of
 $\Gamma_k$
takes the simple form of a renormalization-group-improved 
one-loop equation \cite{exact}
\be
k \frac{\partial}{\partial k} \Gamma_k [\phi]
=\frac{1}{2} {\rm Tr} \left[\left( \Gamma_k^{(2)} [\phi] + R_k \right)^{-1}
 k \frac{\partial}{\partial k} R_k\right].
\label{one} \ee
The trace involves a momentum integration and summation over
internal indices. 
The relevant infrared
properties appear directly in the form of the exact inverse
average propagator $\Gamma_k^{(2)}$, which is the matrix of
second functional derivatives with respect to the fields.
There is always only one momentum integration -- multiloops 
are not needed -- which is, for a suitable cutoff function
$R_k(q^2)$ 
(with $R_k(0) \sim k^2$, $R_k(q^2 \to \infty) \sim e^{-q^2/k^2}$),
both infrared and ultraviolet finite.

The flow equation (\ref{one}) is a functional differential
equation, and an approximate solution requires a truncation. 
Our truncation is the lowest order in a systematic derivative
expansion of $\Gamma_k$ \cite{indices,morris}
\be
\Gamma_k = \int d^3 x \left\{
U_k(\rho_1,\rho_2) + \frac{1}{2} Z_k 
\left( \partial^{\mu} \phi_1 \partial_{\mu} \phi^1 
+ \partial^{\mu} \phi_2 \partial_{\mu} \phi^2 
\right) \right\}.
\label{two} \ee
Here  
$\rho_a = \frac{1}{2} \phi_a \phi^a$ and 
the potential $U_k(\rho_1,\rho_2)$ is symmetric under the interchange
$1 \leftrightarrow 2$.
The wave-function renormalization
is approximated by one $k$-dependent parameter $Z_k$. 
The truncation of the higher derivative terms in the action  
is expected to generate an uncertainty of the order of the anomalous
dimension $\eta$. 
(For $\eta=0$ the kinetic term in the
$k$-dependent inverse propagator must be exactly proportional to 
$q^2$ both for $q^2 \rightarrow 0$ and 
$q^2 \rightarrow \infty$.) For the model we are considering, 
$\eta \simeq 0.035$ and the induced error is small. A similar 
truncation has been employed for the calculation of the 
equation of state for the $O(N)$-symmetric scalar theory \cite{eos}. 
The result is indeed 
in agreement with those obtained through alternative methods
with an accuracy of order $\eta$.

The evolution equation for the potential 
results from the substitution of eq. (\ref{two}) 
in the exact flow equation (\ref{one}). 
The fixed-point structure of the theory 
is more easily identified if
we use the dimensionless renormalized parameters
\be
\rht_a= ~Z_k k^{-1} \rho_a
~~~~~~~~~~~~~~
u_k(\rht_1,\rht_2) = ~k^{-3} U_k(\rhoa,\rhob). 
\label{three}
\ee
The evolution equation for the potential can now be written in
the scale-independent form \cite{stefan}
\be
\frac{\partial}{\partial t} u_k(\rhta,\rhtb) =
-3 u_k +(1 + \eta) (\rht_1 u_1 + \rht_2 u_2) 
- \frac{1}{8 \pi^2} L^3_0(\mt^2_1)
- \frac{1}{8 \pi^2} L^3_0(\mt^2_2),
\label{four} \ee
where $t=\ln(k/\Lx)$.
The anomalous dimension $\eta$ is defined as 
$d \ln Z_k/dt = -\eta$ and can be computed starting from the exact
flow equation \cite{average,indices}. For the model we are considering
and the part of the phase diagram we shall study, it is
constant $\eta \simeq 0.035$ to a good approximation \cite{stefan}. 
The quantities
$\mt^2_{1,2}$ are the eigenvalues of the rescaled mass matrix at the
point $(\rhta,\rhtb)$ 
\be
\mt^2_{1,2} = \frac{1}{2} \left\lbrace 
 u_1 + u_2 +2 u_{11} \rhta + 2 u_{22} \rhtb 
\pm  \left[ 
(u_1 - u_2 +2 u_{11} \rhta - 2 u_{22} \rhtb )^2 
+ 16 u^2_{12} \rhta \rhtb \right]^{\frac{1}{2}} \right\rbrace,
\label{five} \ee
and we have introduced the
notation $u_1 = {\partial u_k}/{\partial \rhta}$,
$u_{12} = {\partial^2 u_k}/{\partial \rhta \partial \rhtb}$, etc. 
The function $L^3_0(w)$, as well as the functions 
$L^3_1(w) = - {d L^3_0(w)}/{d w}$, 
$L^3_{n+1}(w) = - {1}/{n}~{d L^3_n(w)}/{d w}$
for $n \geq 1$ that we shall encounter in the following, are negative 
for all values of $w$. 
Also $|L^3_n(w)|$ are monotonically decreasing for increasing $w$
and introduce a threshold behaviour in the evolution. 
For large values of $\mt^2_a$
the last two terms in eq. (\ref{four}) vanish and the evolution
of $U_k$ stops. The above functions have been extensively 
discussed in refs. \cite{average,indices}.

The initial condition for the integration is provided by the
bare potential, which is identified with the
effective average potential at a very high scale $k=\Lx$. 
We use a bare potential of the form 
\be
 u_{\Lx}(\rhta,\rhtb) = 
\frac{1}{2} {\lx}_{\Lx} \left\{ \left(\rhta-\kx_{\Lx}\right)^2 
+  \left(\rhtb-\kx_{\Lx}\right)^2 \right\} 
+ (1+x_{\Lx})  {\lx}_{\Lx} \rhta \rhtb
\label{twob} \ee
and $Z_{\Lx}=1$.

The phase structure of the theory 
has been discussed in detail in refs. \cite{rudnick,aharony,alford,stefan}.
The phase diagram has three fixed points, which govern the dynamics 
of second-order phase transitions. 
They are located on the critical surface separating the 
phase with symmetry breaking from the symmetric one.
The most stable of them corresponds to a system with 
an increased $O(2)$ symmetry. It can be approached directly from a bare action 
given by eq. (\ref{twob}) with $x_{\Lx}=0$.
The other two 
are Wilson--Fisher fixed points, corresponding to two disconnected 
$Z_2$-symmetric theories. One of them can be approached from a bare action with
$x_{\Lx}=-1$, while the second requires $x_{\Lx}=2$ 
\footnote{
A redefinition of the
fields demonstrates that this choice corresponds to two 
disconnected $Z_2$-symmetric theories \cite{stefan}.}.
Flows that start with $-1 < x_{\Lx} < 2$ and near the critical surface 
eventually lead to the $O(2)$-symmetric fixed point.
For $x_{\Lx} > 2$ or $x_{\Lx} < -1$ the evolution leads
to a region of first order phase transitions. If $x_{\Lx}$ is chosen
slightly larger than 2 or slightly smaller than $-1$ the 
phase transitions are weakly first order. As we demonstrate in the
following, the evolution first approaches
one of the fixed points 
before a second minimum appears in the potential. 

The partial differential 
equation (\ref{four}) with $\eta=0.035$ can be integrated 
numerically through a generalization  
to the case of two fields of the 
algorithms presented in ref. \cite{num} for the case of one background field.
This approach is straightforward but requires 
excessive computer power. An alternative solution relies on 
an approximation that simplifies the form of the potential.
We are interested in
the case where the minima of the potential are located on
the two axes (this requires $x_{\Lx} > 0$).
We concentrate on the $\rhta$ axis and consider the 
region $\rhtb \ll 1$.
The potential can be expanded as
\be
 u_k(\rhta,\rhtb) = v_k(\rhta) + f_k(\rhta) \rhtb + ... 
\label{six} \ee
Substituting the above expression in eq. (\ref{four})
and setting $\rhtb=0$
results in an evolution equation for $v_k(\rhta)$. Notice that 
terms with powers of $\rhtb$ higher than the first 
do not affect the form of this evolution equation. 

We approximate the function $f_k(\rhta)$ by a fourth-order polynomial
in $\rhta$. This is expected to be a good approximation as 
the physical behaviour of interest (scale-invariant 
form of the potential, appearance of a minimum at the origin) is observed 
for $\rhta < 1$ (see fig. 1). 
Near the origin we have 
\be
f_k(\rhta) \simeq m^2_s + \left( 1+x_s \right) \lx_s \rhta,
\label{f1} \ee
with
$m_s^2 = v_1(\rhta=0)$, $\lx_s = v_{11}(\rhta=0)$. 
The form of the first term in the right-hand side is
imposed by the $1 \leftrightarrow 2$ symmetry.
The evolution of the 
$k$-dependent term $x_s$ can be calculated from a 
truncated form of the  
evolution equation at $\rhta=\rhtb=0$ \cite{stefan}
\be
\frac{d x_s}{d t} = \frac{1}{8 \pi^2} (x_s+1) x_s (x_s-2) \lx_s 
L^3_2(m_s^2).
\label{seven1} \ee
Near the minimum of the potential at $\rhta=\kx \not= 0$
we have 
\be
f_k(\rhta) \simeq x_0 \lx_0 \kx 
+ \left( 1+x_0 \right) \lx_0 \left( \rhta-\kx \right),
\label{f2} \ee
with
$\lx_0 = v_{11}(\rhta=\kx)$. 
Again the form of the first term in the right-hand side is
imposed by the $1 \leftrightarrow 2$ symmetry.
The evolution of the 
$k$-dependent term $x_0$ can be calculated from a 
truncated form of the  
evolution equation at $\rhta=\kx$, $\rhtb=0$ \cite{stefan}
\be
\frac{d x_0}{d t} = \frac{1}{8 \pi^2} \frac{6}{\kx}
\frac{x_0+\frac{x_0^2}{4}}{1-\frac{x_0}{2}}
\left\{ L^3_1(2\lx_0\kx)-L^3_1(x_0\lx_0\kx) \right\}
+\frac{1}{8 \pi^2} x_0 \lx_0 
\left\{9 L^3_2(2\lx_0\kx) + \left( 1+x_0 \right)^2 
L^3_2(x_0\lx_0\kx) \right\}.
\label{seven2} \ee
Matching the two expansions of eqs. (\ref{f1}) and (\ref{f2}) 
fixes the coefficients of the fourth-order polynomial.
We shall check the validity of 
this approximation in the following.

The numerical integration of the partial differential equation 
(\ref{four})
is now feasible, because we have only two independent variables
($t,\rhta$) for the function $v_k$ (as well as the 
$k$-dependent parameters $x_s$, $x_0$). The expected numerical accuracy 
of the integration is estimated to be of order 3\% \cite{num}.

We are interested in the region $x \ge 2$. 
The region $x \le -1$ can be mapped on it through a redefinition
of the fields \cite{stefan}.
In fig. 1 we present 
the evolution of
$v_1 = d v_k / d \rhta$, starting at a very high scale $k=\Lx$
with a bare potential 
given by eq. (\ref{twob}).
All dimensionful quantities are normalized 
with respect to $\Lx$. The initial 
coupling $\lx_{\Lx}$ is chosen arbitrarily, while
the minimum $\kx_\Lx$ of the bare potential is
taken very close to the critical value $\kx_{cr}$ 
that separates the phase with symmetry breaking from the symmetric one.
For $|\dkl| = |\kx_\Lx - \kx_{cr}| \ll 1$ the system spends a long ``time''
$t$ of its evolution on the critical surface separating the two phases.
The initial value $x_{\Lx}$ (which determines the
initial conditions $x_s(\Lx)=x_0(\Lx)=x_{\Lx}$ 
for eqs. (\ref{seven1}), (\ref{seven2})) 
is taken slightly larger than the fixed-point value 2.
After the initial evolution (dotted lines) the potential settles down
near the Wilson--Fisher fixed point (solid lines). The fixed point 
of eqs. (\ref{seven1}), (\ref{seven2}) at
$x_s=x_0=2$ is repulsive
\footnote{
Despite the presence of $1-{x_0}/{2}$ in the denominator in the
right-hand side of eq. (\ref{seven1}), this equation has a fixed 
point at $x_0=2$.
}, and $x_s$, $x_0$ eventually evolve towards larger
values. This forces the potential to move away from its scale-independent
form (dashed lines). At some point in the subsequent evolution the 
curvature of the potential at the origin $v_1(\rhta=0)$ becomes
positive. This signals the appearance of a new minimum there, and the 
presence of a fluctuation-driven first-order phase transition. 
The evolution of the potential after it moves away from
the fixed point is
insensitive to the details of the bare potential. It is
uniquely determined by 
$|\dkl| = |\kx_\Lx - \kx_{cr}| \ll 1$ and $\dkx = x_\Lx-2 \ll 1$, 
and, therefore, displays
{\bf universal} behaviour \cite{coarse}. 

In the limit $k \rightarrow 0$ and in the convex regions,
the rescaled potential $u_k$ 
grows and eventually diverges in such a way 
that $U_k$ becomes asymptotically
constant, equal to the effective potential $U=U_0$. This is apparent in
fig. 2, where the potential along the $\rho_1$ axis is plotted. 
The evolution of the non-convex part
of the potential (between the two minima)
is related to the issue of 
the convexity of the effective potential. This part should become flat for
$k \rightarrow 0$ \cite{convex,coarse}. 
The approach to convexity is apparent in fig. 2, 
even though we have not followed 
the evolution all the way to $k=0$.
The nucleation rate during the decay of the unstable
minimum can be calculated in terms of the potential at an
appropriate coarse-graining scale.
It is exponentially suppressed by the free energy of the dominant 
tunnelling configuration.
The scale $k$ must be chosen so that the pre-exponential
factor, arising from fluctuations around the dominant configuration,
is small. An appropriate scale $k_f$ for the 
definition of a universal coarse-grained potential is 
determined by \cite{coarse}
\be
\frac{k^2_f-\left|\left(d^2U_{k_f}/d\phi_1^2\right)_{\phi_1=\phi_{1max}}
\right|}{k^2_f} = 0.01,
\label{lala} \ee
where $\phi_{1max}$ corresponds to the top of the barrier.
The scale at which we have stopped the evolution 
in fig. 2 is near $k_f$.

The relative magnitude of 
$|\dkl| = |\kx_\Lx - \kx_{cr}| \ll 1$ and $\dkx = x_\Lx-2 \ll 1$ 
results in different
types of evolution. For the type of behaviour depicted in figs. 1 and 2 
one must
take $|\dkl| \ll \dxl$. In the opposite limit, $|\dkl| \gg \dxl$,
the system leaves the critical surface before $x_s$, $x_0$ evolve 
away from the fixed-point value 2. As a result, a second minimum never
appears at the origin. Instead, the only minimum of the effective
potential $U_0$ is located
either at zero (symmetric phase) or away from it (phase with symmetry
breaking), depending on the sign of $\dkl$. 
The resulting phase transition is second
order and occurs for $\dkl=0$. 

We are interested in the universal ratio of susceptibilities 
$\chi_+/\chi_-$ on either
side of the phase transition. This ratio depends on the value
of $|\dkl| / \dkx$. For $|\dkl| / \dkx \gg 1$ we have 
$x_s(k),x_0(k)  \simeq 2$ during
the whole evolution. The universal quantities characterizing the
second-order phase transition are determined by the Wilson--Fisher
fixed point. We calculate 
$\chi_+/\chi_-$ by integrating the evolution equation 
and evaluating 
$d^2 U_0/d \phi_1^2 = \chi^{-1}$ at the minimum, 
for $\dkl = \mp \epsilon$ with
$\epsilon \ll 1$.
We obtain $\chi_+/\chi_- = 4.1$. This value should be compared with 
the value $\chi_+/\chi_- = 4.3$, calculated with the same method 
for the one-field, $Z_2$-symmetric theory (the Ising model)
without any polynomical approximations for the form of the
potential \cite{eos}. 
The difference
is due to the approximation of the function $f_k(\rhta)$ 
in eq. (\ref{six}) by a fourth-order polynomial.
An error of this magnitude is typically induced by 
the truncation of the potential to a polynomial form \cite{indices,eos}. 
We have verified this conclusion by approximating the function
$f_k(\rhta)$ by eq. (\ref{f1}) without using eq. (\ref{f2}).
This cruder approximation
results in $\chi_+/\chi_- = 4.0$.

The evaluation of the same quantity through the $\epsilon$-expansion or
an expansion at fixed dimension gives 
$\chi_+/\chi_- = 4.8(3)$ \cite{zinn}, whereas 
experimental information gives $\chi_+/\chi_- = 4.3(3)$
\cite{zinn}.
The difference with the results of our method 
is due to the omission of higher derivative terms 
in the effective average action of eq. (\ref{two}).
This confirms our expectation, discussed earlier, that 
the error induced by this approximation is
determined by the anomalous dimension, which is small
for our model ($\eta \simeq 0.035$). 

For $|\dkl| / \dkx \ll 1$ the potential develops a second minimum at the
origin during
the later stages of the evolution. The phase transition is approached by 
fine tuning $\dkl$, so that the two minima have equal depth for $k=0$. 
In fig. 3 we plot the difference in energy density $\Delta U$
between the minimum 
away from the origin and the one at the origin, as a function of the 
scale $k$, for four values of $\dkl$. The numerical 
integration of the evolution
equation is difficult for $k \rightarrow 0$ because of the 
singularity structure of $L^3_n(w)$ \cite{convex,coarse}. As a result the
curves of fig. 3 must be extrapolated to $k=0$. 
Line (a) corresponds to a system in the symmetric phase, line
(d) to one in the phase with symmetry breaking. 
Lines (b) and (c) correspond to a system very close to the phase transition. 
The evolution of 
$\chi_+/\chi_-$ for the same parameters is depicted in fig. 4. 

The extrapolated value of $\chi_+/\chi_-$ for $k=0$ near the 
phase transition is expected to lie in the interval (1.5, 2). 
We have used numerical fits of various curves for $\Delta U$ and 
$\chi_+/\chi_-$ in the vicinity of the 
phase transition in order to perform the extrapolation to $k=0$. 
The expected value for $\chi_+/\chi_-$ is 1.7. 

The main sources of error in our calculation 
are the derivative expansion of eq. (\ref{two}) and the approximation of 
$f_k(\rhta)$ in eq. (\ref{six}) by a fourth-order polynomial
in $\rhta$. We argued earlier that the error induced by the
derivative expansion is related to the small anomalous dimension
$\eta \simeq 0.035$.
We also checked above the error induced by the polynomial 
appoximation of the potential. For the second-order phase transition,
two polynomial approximations give 
$\chi_+/\chi_-=4.0$ and 4.1, while no polynomial approximation gives
$\chi_+/\chi_-=4.3$ in the order of the derivative expansion we are
working \cite{eos}. The numerical integration and the  
extrapolation used in the calculation of $\chi_+/\chi_-$ induce smaller
errors. The $\epsilon$-expansion or
an expansion at fixed dimension gives 
$\chi_+/\chi_- = 4.8(3)$ \cite{zinn}, whereas 
experimental information gives $\chi_+/\chi_- = 4.3(3)$
\cite{zinn}.
In order to be conservative, we use as a maximum total error 
the difference between
our result $\chi_+/\chi_-=4.1$ for the case of the second-order
phase transition
and the central value 
$\chi_+/\chi_-=4.8$ of the result
from the $\epsilon$-expansion.

Our result 
\be
\chi_+/\chi_-=1.7(7)
\label{resu} \ee
is depicted by a horizontal band in
fig. 4.
It is in good agreement with the 
predictions of the
$\ex$-expansion ($\chi_+/\chi_-=2.0, 2.9, 2.3$ 
for the first three orders)
\cite{arnolde,arnolds}.
It disagrees, however, with the lattice result presented in refs. 
\cite{arnoldl,arnolds} ($\chi_+/\chi_-=4.1(5)$). The values 
favoured by the lattice calculation correspond
to a theory well into the phase with symmetry breaking in our calculation
(line (d) in figs. 3 and 4). However, it should be noted that, according to
the authors of ref. \cite{arnolds}, the error of the lattice result
``should be taken with a grain of salt''. 

In summary, we presented a method that can provide a quantitative 
description of the universal behaviour near 
weakly first-order phase transitions.
It is based on the calculation of a coarse-grained free
energy through an exact flow equation. Fixed points in the evolution, 
the appearance of new minima in the potential, and the universal 
properties of the
resulting fluctuation-driven first-order phase transitions 
can be studied in detail. We calculated 
the universal ratio of susceptibilities 
$\chi_+/\chi_-$ on either 
side of the first-order phase transition
in a two-scalar model .

\newpage
\pagestyle{empty}

\begin{figure}
\psfig{figure=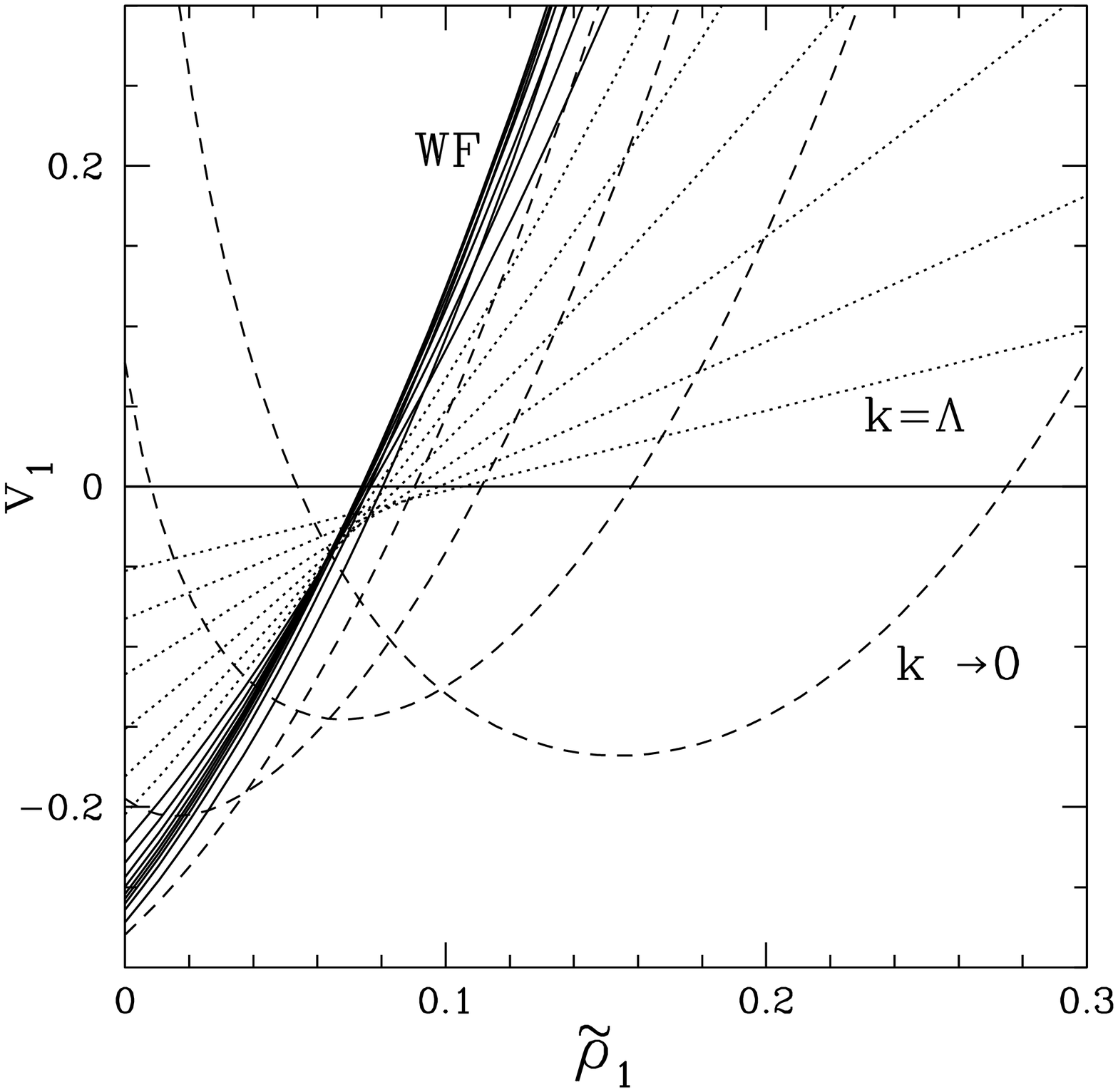,height=16.0cm}
Fig. 1: The derivative of the rescaled 
potential along the $\rho_1$ axis, as the coarse-graining scale
is lowered. 
 
\end{figure}

\begin{figure}
\psfig{figure=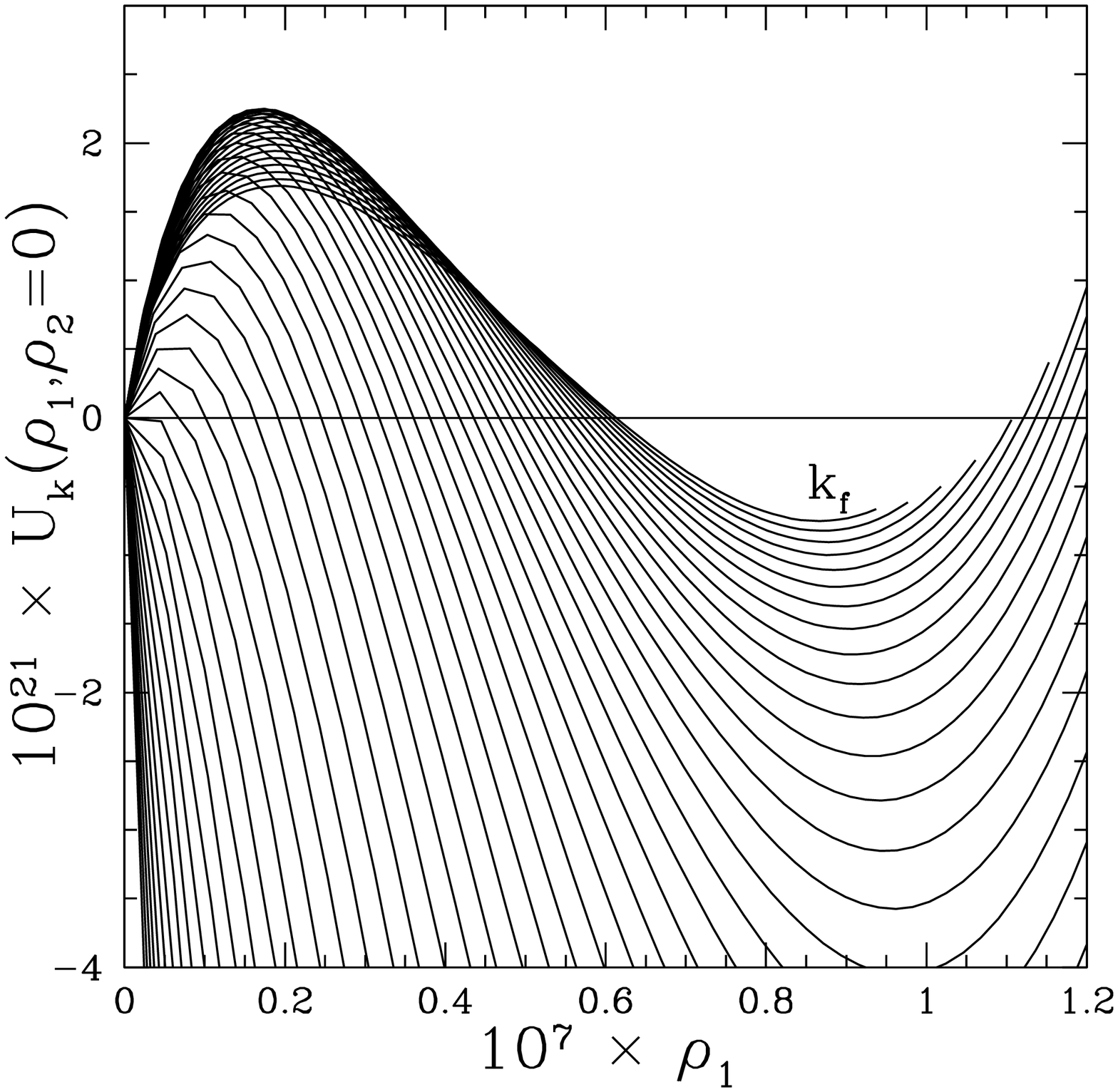,height=16.0cm}
Fig. 2: The potential along the $\rho_1$ axis, as the 
coarse-graining scale is lowered.
 
\end{figure}

\begin{figure}
\psfig{figure=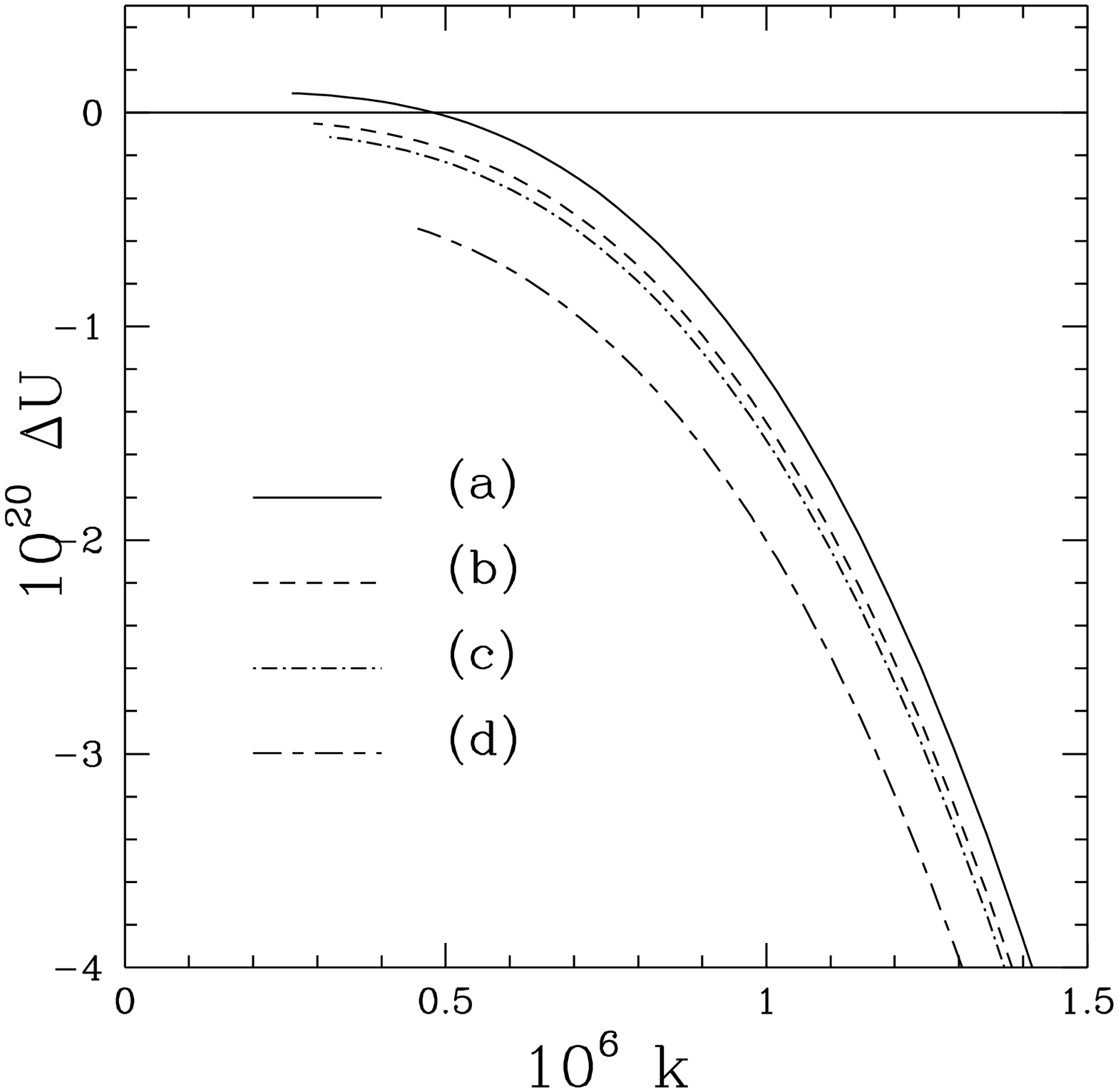,height=16.0cm}
Fig. 3: The difference in energy density between the two minima of the 
potential, as a function of the coarse-graining scale, near the 
phase transition. 
Line (a) corresponds to a system in the symmetric phase, line
(d) to one in the phase with symmetry breaking. 
Lines (b) and (c) correspond to a system very close to the phase transition.

\end{figure}

\begin{figure}
\psfig{figure=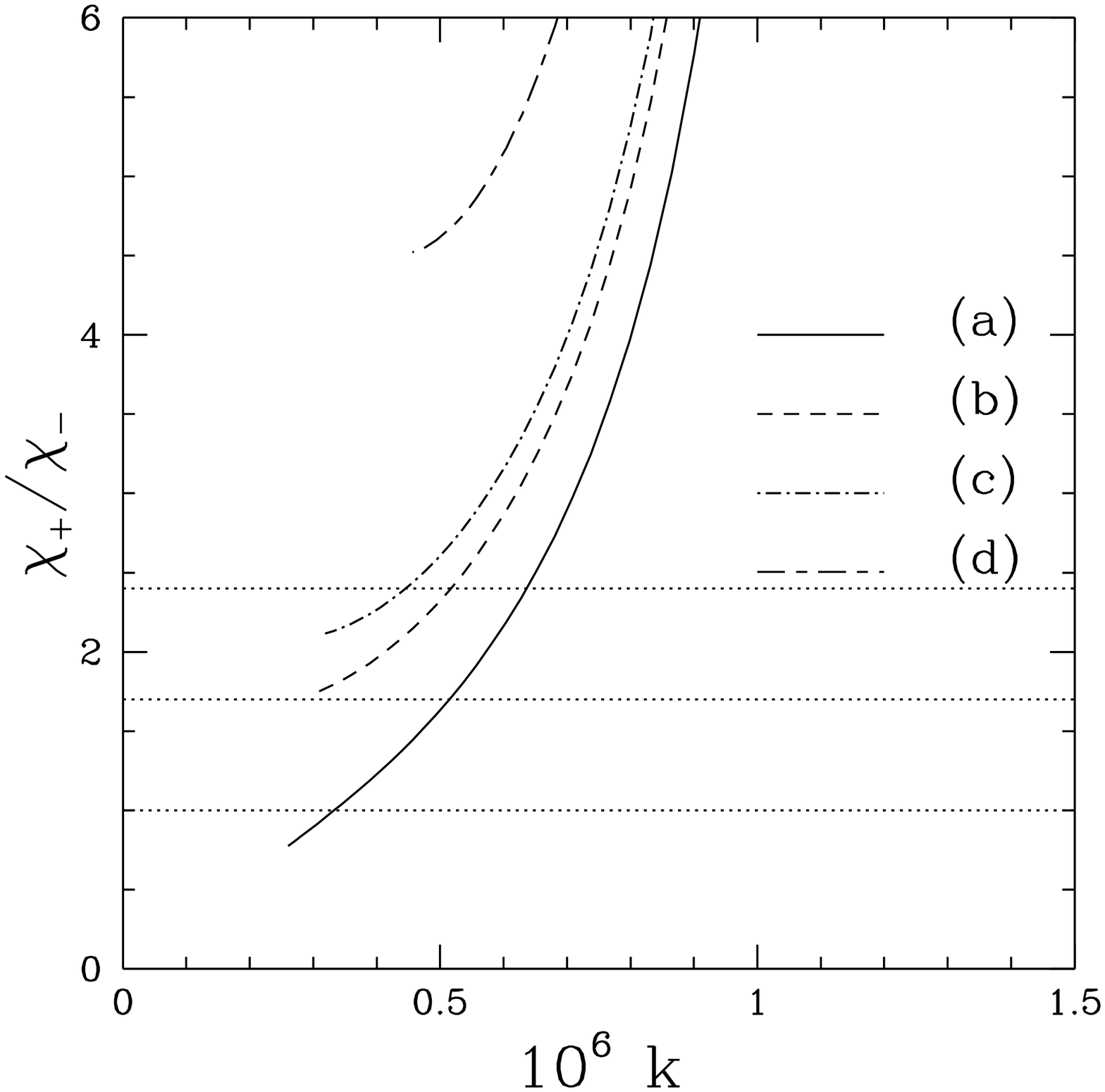,height=16.0cm}
Fig. 4: 
Analogous to fig. 3 for the ratio of susceptibilities $\chi_+/\chi_-$. 
 
\end{figure}


\newpage

\begin{thebibliography}{99}

\bibitem{review}
K. Kajantie, M. Laine, K. Rummukainen and M. Shaposhnikov, 
Phys. Rev. Lett. {\bf 77}, 2887 (1996);
V.A. Rubakov and M.E. Shaposhnikov, 
preprint CERN-TH/96-13, hep-ph/9603208; 
K. Jansen, 
preprint DESY-95-169, hep-lat/9509018.

\bibitem{me}
N. Tetradis, Nucl. Phys. B {\bf 488}, 92 (1997).

\bibitem{rudnick}
J. Rudnick, Phys. Rev. B {\bf 18}, 1406 (1978).

\bibitem{aharony}
A. Aharony, in: Phase Transitions and Critical Phenomena, vol. 6,
eds. C. Domb and M.S. Green, Academic Press (1976);
D.J. Amit, Field Theory, the Renormalization Group, and Critical Phenomena,
World Scientific (1984).

\bibitem{transition}
N. Tetradis and C. Wetterich, Nucl. Phys. B {\bf 398}, 659 (1993).

\bibitem{alford}
M. Alford and J. March-Russell, Nucl. Phys. B {\bf 417}, 527 (1994).

\bibitem{stefan}
S. Bornholdt, N. Tetradis and C. Wetterich,
Phys. Lett. B {\bf 348}, 89 (1995);
Phys. Rev. D {\bf 53}, 4552 (1996);
S. Bornholdt, P. B\"uttner,  N. Tetradis and C. Wetterich,
preprint CERN-TH/96-67, cond-mat/9603129.

\bibitem{arnolde}
P. Arnold and L.G. Yaffe, Phys. Rev. D {\bf 55}, 7760 (1997);
P. Arnold and Y. Zhang, Phys. Rev. D {\bf 55}, 7776 (1997).

\bibitem{arnoldl}
P. Arnold and Y. Zhang, Nucl. Phys. B {\bf 501}, 803 (1997).

\bibitem{arnolds}
P. Arnold, S.R. Sharpe, L.G. Yaffe and Y. Zhang, 
Phys. Rev. Lett. {\bf 78}, 2062 (1997).

\bibitem{wilson}
K.G. Wilson, Phys. Rev. B {\bf 4}, 3174 and 3184 (1971);
K.G. Wilson and I.G. Kogut, Phys. Rep. {\bf 12}, 75 (1974);
F.J. Wegner, in: Phase Transitions and Critical Phenomena, vol. 6,
eds. C. Domb and M.S. Green, Academic Press (1976). 

\bibitem{average}
C. Wetterich, Nucl. Phys. B {\bf 352}, 529 (1991);
Z. Phys. C {\bf 57}, 451 (1993); ibid. C {\bf 60}, 461 (1993);
Phys. Lett. B {\bf 301}, 90 (1993).

\bibitem{exact}
C. Wetterich, Phys. Lett. B {\bf 301}, 90 (1993).

\bibitem{indices}
N. Tetradis and C. Wetterich, Nucl Phys. B {\bf 422}, 541 (1994).

\bibitem{morris}
T. Morris, Nucl. Phys. B {\bf 495}, 477 (1997).

\bibitem{eos}
J. Berges, N. Tetradis and
C. Wetterich, 
Phys. Rev. Lett. {\bf 77}, 873 (1996).

\bibitem{num}
J. Adams, J. Berges, S. Bornholdt, F. Freire, N. Tetradis and
C. Wetterich, 
Mod. Phys. Lett. A {\bf 10}, 2367 (1995). 

\bibitem{convex}
A. Ringwald and C. Wetterich, Nucl. Phys. B {\bf 334}, 506 (1990);
N. Tetradis and C. Wetterich, Nucl. Phys. B {\bf 383}, 197 (1992).

\bibitem{coarse}
J. Berges and C. Wetterich, 
Nucl. Phys. B {\bf 487}, 675 (1997);
J. Berges, N. Tetradis and C. Wetterich,
Phys. Lett. B {\bf 393}, 387 (1997).

\bibitem{zinn}
J. Zinn-Justin, Quantum Field Theory and Critical Phenomena, Oxford Science
Publications (1989).










\end{thebibliography}
\end{document}